\documentclass[11pt]{article}

\usepackage[utf8]{inputenc}
\usepackage{authblk}      
\usepackage{natbib}
\usepackage[most]{tcolorbox}
\usepackage{float}
\usepackage{graphicx}
\usepackage{caption}
\usepackage{url}
\usepackage{natbib}

\title{Measuring ESG Risk in Supply Networks}
\author[1]{Rudy Arthur\thanks{Corresponding author: \texttt{R.Arthur@exeter.ac.uk}}}
\author[1,2]{Guillherme Machado}
\affil[1]{Department of Computer Science, University of Exeter}
\affil[2]{Department of Physics, Universidade de Aveiro}

\usepackage{xcolor}

\date{\today}

\begin{document}

\maketitle

\begin{abstract}	
 \noindent Environmental, Social and Governance (ESG) rating is a way for investors to prioritise investments in companies with good corporate behaviour. However, ESG ratings are vulnerable to greenwashing in a number of ways. In this paper we study the effect that trade with badly rated companies has on a target company's own rating. To do this we introduce a measurement framework, generalising PageRank and Alpha Centrality, which allows tuning of aggregation and path counting approaches to resist greenwashing and reflect the rater's opinions and preferences for harm accumulation. These metrics allow updating of the target's ESG rating, identification of influential neighbours and assessment of vulnerability of the target to bad behaviour in their supply network. We study these metrics on synthetic ESG interaction networks as well as a real inter-company network and the international trade network.
\end{abstract}

\section{Introduction}\label{sec:introduction}

Rising to prominence in 2004 \cite{UN2004WhoCaresWins}, ESG, \textbf{E}nvironmental, \textbf{S}ocial and \textbf{G}overnance, is an investing principle building on prior ideas of corporate social responsibility (CSR) \cite{sheehy2015defining}. ESG refers to both the way that investors evaluate corporate behaviour according to ethical principles, as well as an investing strategy used to assess future financial performance \cite{li2021esg}. Total assets under management in so-called responsible investment funds, which prioritise ESG rating over other metrics, are in the tens of trillions according to industry reports \cite{GSIA2025_GSIR2024}, although precise numbers vary. 

ESG relies on accurate analysis of corporate behaviour. Ratings are provided by a number of companies and funds, with the most prominent in the academic literature being the Thomson Reuters ASSET4 database \cite{de2022corporate} and the MSCI KLD 400 Social Index \cite{eccles2020social} due to their accessibility, although many other ratings and ESG funds exist. Since ratings are sold as products to investors, they tend not to be freely available and their methodology not fully transparent, however the broad idea is clear and described in various industry documents e.g. \cite{MSCI2024_ESGRatingsMethodology}. Calculating a company's ESG rating involves the analysis of the company's Environmental, Social and Governance issues, which can be as diverse as water use, data privacy or board diversity with a weighted combination determining the final rating. 

With large amounts of investment capital at stake, companies have responded by trying to improve their ESG ratings, which is the intention of the architects of ESG. However, such attempts are not always honest and corporate greenwashing, misleading information about ESG performance \citep{delmas2011drivers}, is an issue. Greenwashing can involve selective disclosure of information \cite{lyon2011greenwash} or using past ethical behaviour to cover for current unethical behaviour \cite{parguel2011sustainability}. \cite{kathan2025you} find that good ESG ratings are only moderately correlated with good environmental performance and positively correlated with the number of greenwashing cases reported for that company. 

\begin{figure}
    \centering
    \includegraphics[width=0.95\linewidth]{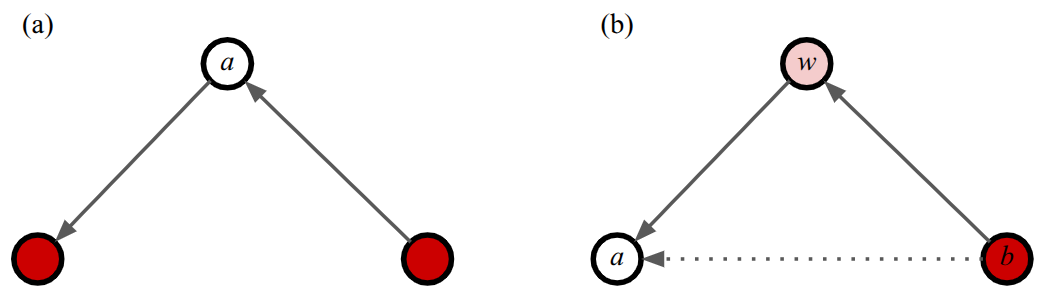}
    \caption{(a) Shows an ostensibly good company $a$, with a good ESG rating, being supplied by and supplying badly rated companies. (b) Shows a company with a bad ESG rating, $b$ effectively trading with $a$ through an intermediary $w$.}
    \label{fig:1}
\end{figure}
Corporate networks create opportunities, deliberate or not, for greenwashing. Figure \ref{fig:1}(a) shows a simple example: company $a$ has good ESG metrics, but is supplied by a badly rated company (e.g. a highly polluting mine) and sells to badly rated customers (e.g. aggressive militaries). It is reasonable to suppose ethical investors would be concerned by such associations. Failing to incorporate them into the ESG rating should be considered a form of greenwashing. Figure \ref{fig:1}(b) shows a slightly more subtle example, where company $b$ with low ESG rating, can form a `clean' subsidary or wholesaler, $w$, which launders the reputation of company $b$ allowing other companies to invest and trade with the subsidiary without the reputational damage that might come from trading with $b$ directly. Many similar scenarios can be envisaged and need not be deliberate to be a form of greenwashing.

Rating agencies do incorporate supply chain effects \cite{berg2022aggregate}, however, as with all aspects of ESG rating, the methods are opaque. In the academic literature some studies exist on the interaction of corporate networks with ESG ratings. \cite{xie2023can, yang2024conformity} suggest peer effects operate, where network attention enhances the ESG performance of companies in a network. \cite{ang2023predicting} uses this operationally, using company networks to \emph{predict} ESG scores. Most of this literature uses regression based approaches i.e. modelling ESG score as a function of various factors like network attention. In this paper we use tools from network science which give us a more detailed picture of how ESG scores are affected by the structure of the corporate network. This is, to our knowledge, a novel approach with the closest work we can locate from Ioannidis et. al. \cite{ioannidis2022correlations} who use network science to study the network produced from correlations between different ESG components, rather than the company network itself.

It is important to note that unlike credit rating, ESG ratings from different issuing agencies are only moderately correlated \cite{chatterji2016ratings, berg2022aggregate}. This reflects the opinions and intentions of different raters on how to quantify or even define the ESG concept \cite{dorfleitner2015measuring, eccles2018exploring}. This ambiguity is somewhat inevitable. An ESG rating attempts to combine disparate aspects of corporate behaviour, from polluting a river to worker harassment, into a single score and this involves many value judgements which have no unambiguous resolution. Our view in this paper is that the ambiguity in ESG scores is a design constraint. Different raters want to evaluate companies in different ways with different priorities. A useful `rating tool' should allow for diverse opinions on what matters for ESG.

In this paper we develop metrics that allow us to update a company's ESG rating based on their network. To do so we present a deconstruction and generalisation of some well known tools from network science to build a flexible framework for network ESG analysis. This  allows an investor or rater to specify how different scenarios should be scored. We provide guidance on how robust or vulnerable different approaches are to greenwashing or obfuscation patterns. Ultimately the aim is that if network effects are incorporated into ESG ratings, then the ratings will become strongly interdependent. This would lead to community enforcement of good practices, with badly rated suppliers or customers excluded out of self-interest, enhancing the peer effects described in previous research \cite{xie2023can, yang2024conformity}, realising the aims of ESG proponents.

In Section \ref{sec:methods} we introduce our metrics, showing how they come from generalising well known metrics like PageRank to cover other forms of score aggregation and path counting. In Section \ref{sec:small} we analyse some networks designed to show the metrics in simplified settings to build intuition. In Section \ref{sec:realdata} we cover two case studies on real networks, a company network and the international trade network. We conclude in Section \ref{sec:conclusion}.

\section{Network ESG Metrics}\label{sec:methods}

\begin{figure}
    \centering
    \includegraphics[width=0.55\linewidth]{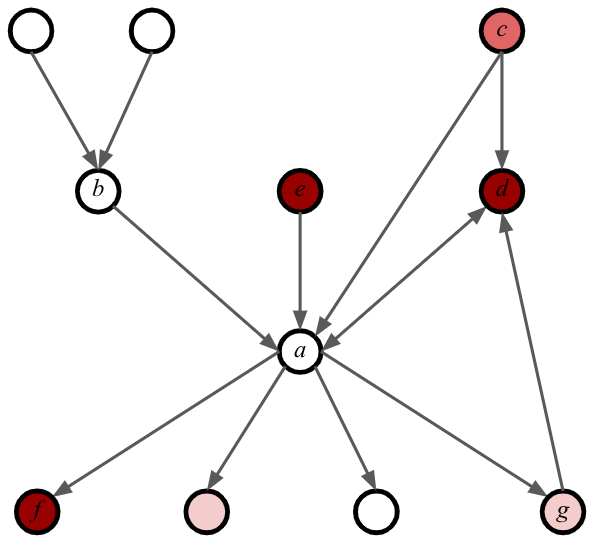}
    \caption{An example supply network. We will mostly focus on the target company $a$. Colours represent ESG ratings, the more red, the worse the rating.}
    \label{fig:2}
\end{figure}

 Different rating agencies use different scales for ESG scores. For example, MSCI uses grades CCC to AAA and Sustainalytics uses values between 0 and 100 \cite{eccles2018exploring}. We will find it convenient to transform the ESG rating for company $a$ into a score $h(a)$ taking values between $0$ and $100$, where a company with a perfect ESG rating would receive a score of $0$ and the worst possible rating is $100$. As a shorthand we will refer to $h(a)$ as `harm', since worse ESG rating corresponds to a higher $h(a)$ caused by company $a$ doing some action counter to ESG principles and therefore causing some kind of harm to people or nature. For concreteness we could take $h(a) = (100 - \text{Sustainalytics})$ or divide the range of MSCI grades into bins with AAA corresponding to $0$ and $CCC$ to $100$.

Consider Figure \ref{fig:2}, which represents a (fictional) company, $a$, together with its supply network. Companies can enter into many different relations with one another like (complete or partial) ownership or partnership. To keep the discussion simple, we will refer to all relations as `supply', although when we come to analyse real company networks in Section \ref{sec:ekodata} we will consider other types. The colours represent the ESG rating of each company \emph{without considering network effects}. Formally, the companies are represented by nodes, $a$, with node values $h(a)$, and are connected by directed edges $(a,b)$. Our goal is to construct a metric $H(a)$ which we will call `network harm'. This should incorporate information about the ESG ratings of the companies with which $a$ interacts. Intuitively, in Figure \ref{fig:2}, $a$ should have a high $H(a)$ value, since it is supplied by two badly rated companies, $e$ and $d$ and supplies the badly rated $f$ and $d$. If $b$ was the target company, $H(b)$ would be lower since $b$ is supplied by companies with good ratings and doesn't directly supply bad ones, though we might want to incorporate the `second order' supply of $b$ to $f$ via $a$.

In the rest of the paper we will develop a variety of metrics which quantify different aspects of harm transmission or `guilt by association' to construct a number of network ratings $H_1(a), H_2(a), \ldots$. These network scores are intended to be weighted and combined with other data to give the final ESG score, e.g. 
$$
{\cal H}(a) = w_h h(a) + w_{H_1} H_1(a)  + w_{H_2} H_2(a)  + \ldots
$$
The weighted sum can be replaced by more sophisticated rules, but the linear approach can accurately approximate most schemes \cite{berg2022aggregate}. ${\cal H}$ can then be transformed into an ESG rating in any scale. The combination and weighting will depend on the rater's priorities, so we don't consider this and focus here exclusively on how to calculate the network score $H(a)$.

\subsection{PageRank and Alpha-Centrality}\label{sec:pagerank}

Although there are a number of network metrics which incorporate node scores \cite{abbasi2013hybrid, singh2020node, de2021centrality, arthur2025correlation} by far the most common are Alpha-Centrality \cite{bonacich2001eigenvector} (which is closely related to Katz centrality \cite{katz1953new}) and PageRank with personalization \cite{haveliwala2002topic}. Alpha-Centrality is defined by
\begin{equation}
    x = \alpha Ax + (1-\alpha) \beta
\end{equation}
where $x$ is the vector of centrality scores, $A$ is the adjacency matrix of the network, $\beta$ is a vector of node scores and $\alpha<1$ is a factor used to balance network effects, represented by the first term, with intrinsic scores, represented by the $\beta$ term. $\alpha < \frac{1}{\lambda_{max}(A)}$ is required for convergence. PageRank with personalization is defined similarly
\begin{equation}
    x = \alpha Px + (1-\alpha) \beta
\end{equation}
Where 
$$
P_{ij} = A_{ij}/k_j^{out}
$$
discounts `link hubs', that is pages which link to many others but are not themselves the targets of many links. PageRank is defined for any $\alpha < 1$, since $P$, as a stochastic matrix, has better convergence properties than $A$. 

Assuming $\alpha$ is sufficiently small we can formally solve for Alpha-Centrality
\begin{equation*}
    x = (1-\alpha) (\mathbf{1} - \alpha A)^{-1} \beta
\end{equation*}
with a similar equation replacing $A$ with $P$ for PageRank. Alpha-Centrality and PageRank can also be constructed as a sum over `levels'. That is, summing over node values at the origin of paths of length 1, paths of length 2 and so on. To see this for Alpha-Centrality, the sum of the intrinsic scores of the direct neighbours of node $i$ is
$$
\sum_{j} A_{ij} \beta_j
$$
The sum of scores of the two-step neighbours is
$$
\sum_{j} (A^2)_{ij} \beta_j
$$
$(A^m)_{ij}$ counts paths of length $m$ between $i$ and $j$. Weighting paths of length $m$ by a factor $\alpha^m$, we have the sum over all levels
\begin{equation*}
    x'_i = \sum_{m = 0}^{\infty} \sum_j \alpha^m (A^m)_{ij} \beta_j
\end{equation*}
Performing the sum gives
\begin{equation}
    x' = (\mathbf{1} - \alpha A)^{-1} \beta
\end{equation}
so $(1-\alpha) x' = x$ and we can view Alpha-Centrality as a weighted sum of contributions from nodes at each level. An identical calculation can be performed for PageRank with $A$ replaced by $P$.

Alpha-Centrality or PageRank in their usual form are not ideal metrics for computing ESG scores. The first reason is that they sum. ESG ratings are not cumulative, like say, total $CO_2$ production which is studied in Life Cycle Analysis \cite{guinee2011life}. For ESG analysis, the meaning of adding an A rating to a C rating is not obvious. By converting to numerical scores we could make Alpha-Centrality or PageRank meaningful in a statistical sense, say, by comparing scores of all companies in the same industry. However this may not be feasible for every use case and we would prefer a metric which can be assessed in isolation.

Alpha-Centrality has the undesirable feature that $\alpha$ must be smaller than the largest eigenvalue of $A$, forcing us to suppress longer paths in order to compute it, potentially hiding bad actors deep in the supply chain. While PageRank converges for any $\alpha < 1$, by using a weight inversely proportional to the out-degree of each node, $\frac{1}{k_j}$, nodes with many outgoing links have their influence suppressed. For ESG rating, buying from a harmful company shouldn't count less simply because it has many other customers. In particular this would seem to advantage primary producers like oil companies and big agribusiness who sell to a many different customers. 

Both these standard measures force us into decisions, longer paths are less important, large producers have their harm reduced, which aren't always desirable from an ESG rating perspective. In line with the diversity of ESG ratings highlighted by \cite{eccles2018exploring} we require more flexibility in our metrics. The two most important aspects are how we aggregate node values and how we count paths. For aggregating we might be happy with average ratings or we might want to take a more stringent view, that a company's network ESG score should be determined by its \emph{worst} relationships. For example in Figure \ref{fig:2}, $a$'s reputation is only likely to be damaged by its supply to the particularly bad customer $f$ and its bad suppliers $d$ and $e$. Figure \ref{fig:2} also raises questions about path counting. Including all paths, as in PageRank or Alpha-Centrality, double counts (or more) some companies. For example, $c$ is directly linked to $a$ and also linked to it via a path of length $2$ through $d$. Should we count $c$ twice? Loops, like $a,g,d$ or even $a,d$, generate infinitely many paths, how should we deal with this?

The next two sections will discuss how we can combine ratings and adapt the `level counting' approach from Alpha-Centrality and PageRank to be more suitable for ESG analysis.

\subsection{Aggregation}

We combine the ESG scores of the direct suppliers of node $a$ using an aggregation function to compute a score we will refer to as the \textbf{upstream harm}
\begin{equation}\label{eqn:agg}
    x_{AGG}(a) =  \text{AGG}_{i \in N^{in}(a)} h(i)
\end{equation}
where $N^{in}(a)$ are the direct suppliers of $a$ and $AGG$ is a function that performs some aggregation operation, like average or maximum, over a multi-set of nodes, returning a scalar value.

For example if we wanted to judge a company by its worst association we would use a $\max$ aggregation function
\begin{equation}
    x_{MAX}(a) = \text{MAX}( \{ h(i) : i \in N^{in}(a) \})
\end{equation}
If we judged a company by the average of its inputs we can use an average aggregation
\begin{equation}
    x_{AVG}(a) = \frac{1 }{|N^{in}(a)|} \sum_{i \in N^{in}(a)} h(i)
\end{equation}

\begin{figure}
    \centering
    \includegraphics[width=0.95\linewidth]{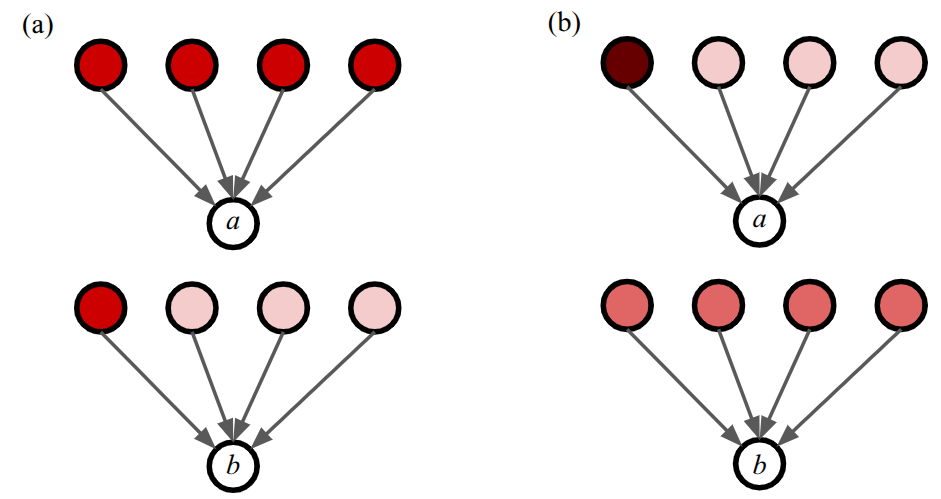}
    \caption{(a) Under max aggregation $x_{MAX}(a) = x_{MAX}(b)$, the behaviour of any company but the worst is irrelevant. (b) Under average aggregation $x_{AVG}(a) = x_{AVG}(b)$, a bad association can be compensated by good ones, giving a neutral effect.}
    \label{fig:3}
\end{figure}
There are a number of things to note about even the simple process of aggregating the harms of direct suppliers. First, as widely understood from the literature on graph machine learning \cite{xu2018powerful}, aggregation loses information. Figure \ref{fig:3} shows some simple patterns which are counter-intuitive for either max or average aggregation. 

Max and average represent two extremes which we can interpolate using TOP-$k$ aggregation
\begin{align}
    x_{TOP-k}(a) &= \frac{1}{|N^{in}(a; k)|} \sum_{i \in N^{in}(a;k)} h_i \\
    N^{in}(a;k) &= \{i : h(i) \text{ is in the top } k\% \text{ of } \{h(j)\}_{j \in N^{in}(a)} \}
\end{align}
We use percentiles so as not to bias against companies with many inputs. As $k$ goes from $0$ to $100$, TOP-$k$ interpolates between max and average. There are many other types of aggregation: variance, softmax, entropy etc. For ESG rating, max and average represent two extremes: max punishes a company for having \emph{any} associations with harmful companies and average punishes a company for having overall harmful associations. TOP-$k$ interpolates between these extremes e.g. TOP-$50$ punishes companies who have more than half of their associations with harmful companies.

The same formula can be applied to the customers of $a$ by replacing $N^{in}(a)$ with $N^{out}(a)$, the set of companies supplied by $a$, giving
\begin{equation}
    \tilde{x}_{AGG}(a) = \text{AGG}_{i \in N^{out}(a)}h_i
\end{equation}
which we call \textbf{downstream harm}, the amount of harm done by the users of $a$'s outputs.

\subsection{Path Counting}

One way for a company to avoid guilt by association is, following Figure \ref{fig:1}(b), to buy from a harmful company through an intermediary or wholesaler. Deliberate or not, such relations obfuscate how a company is connected to harmful actors deeper in its supply chain. For example, if a company assembles components from companies with good ESG ratings, but these suppliers rely on harmful inputs (for example, conflict minerals in electronic devices \cite{fitzpatrick2015conflict}) this should factor into the network rating of the target company. In this way we may start to see companies enforcing standards on suppliers and causing ripples of improvement through supply chain. To do this we need to consider relations beyond first order and incorporate the entire network into the score.

Figure \ref{fig:2} shows a situation which can arise where a company, $c$, is both a first and second order neighbour of $a$. Should $a$ pay a double penalty for its association with $c$? As discussed in Section \ref{sec:pagerank} the standard practice in network science is to sum over all paths and so include two contributions from $c$, with the contribution from the path of length 2 damped by an extra factor of $\alpha$. Figure \ref{fig:2} also has a loop, $a,g,d$. When a loop exists there are paths of arbitrary length, and we end up accumulating harm from the same company infinitely many times. These cycles are what force us to choose small alpha in the case of Alpha-Centrality or to divide by the out-degree in the case of PageRank. 

\begin{figure}
    \centering
    \includegraphics[width=0.95\linewidth]{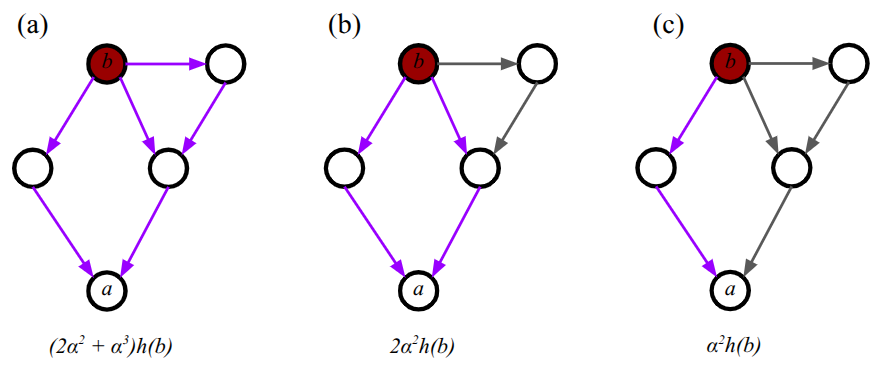}
    \caption{Contribution from paths originating at $b$ to the harm at $a$ using (a) all paths (b) all shortest paths (c) shortest path.}
    \label{fig:4}
\end{figure}
Some alternative path counting approaches are illustrated in Figure \ref{fig:4}. The first counts all \emph{simple} paths, that is, paths without loops, rather than all paths. This eliminates the problem of loops, by definition. The second considers only shortest paths. The last is the most restrictive, considering only shortest paths and including each node only once, at the lowest level at which it appears. In practice, see Section \ref{sec:ekodata}, we are only likely to know the company network to some shallow depth, so the networks are relatively small and computing all the paths from any node to the target is feasible with standard algorithms.

We can write down the contribution from paths of length $m$ as
\begin{equation}
    x_{AGG}^m(a; {\cal P}^m(a) ) =  \text{AGG}_{i \in {\cal  P}^m(a) } h(i)
\end{equation}
Where ${\cal P}^m(a)$ is the multi-set of all the nodes originating a path of length $m$ to $a$, under whatever path counting scheme we choose. For example,
\begin{align*}
     {\cal A}^{m}(a) &= 
   \text{multi-set of nodes beginning any path of length } m \text{ to } a \\
     {\cal C}^{m}(a) &= 
   \text{multi-set of nodes beginning simple paths of length } m \text{ to } a \\
     {\cal S}^{m}(a) &= 
   \text{multi-set of nodes beginning shortest paths of length } m \text{ to } a 
\end{align*}
and to count just one shortest path we use
\begin{equation*}
     \bar{ \cal S}^{m}(a) = \text{set of nodes beginning simple paths of length } m \text{ to } a 
\end{equation*}
For $m=1$, ${\cal A, C, S}$ and $\bar{\cal S}$ are equivalent and for $m>1$ they are equivalent in the special case of a tree (see Figure \ref{fig:5}). For a general network they are different. For downstream harm (judging a company by who it sells to) we have
\begin{equation}
    \tilde{x}_{AGG}^m( a; \tilde{{\cal P}}^m(a) ) =  \text{AGG}_{i \in \tilde{{\cal P}}^m(a) } h(i)
\end{equation}
where $\tilde{{\cal P}}^m(a)$ is now a multiset of nodes which are the \emph{endpoints} of paths of length $m$ starting from $a$.

As with aggregation, there is no one-size-fits-all solution to path counting. Depending on how we envisage harm transmission we may use different approaches. Harmful companies can taint multiple levels of the supply network and the target company is effectively supporting their activities multiple times. For some raters counting all paths may be appropriate, others might consider this overly punitive. The right approach depends on the needs of the user and we will study different choices.

\subsection{Network Harm}

We define the upstream \textbf{network harm} as
\begin{equation}
    H_{AGG', AGG}(a; {\cal P}) = \text{AGG'}_m^{m_{max}} \,\, \alpha^{m-1} x_{AGG}^m( a; {\cal P}^m(a) )
\end{equation}
Where $m_{max}$ is the length of the longest path to be considered. This definition aggregates the upstream harm $x^m_{AGG}$ at each level $m$ and encompasses PageRank, Alpha-Centrality and a large number of other metrics. Note we do not need to use the same aggregation function to aggregate \emph{over} the levels as we do \emph{within} the levels. As before, we can also define the downstream network harm replacing $x$ with $\tilde{x}$.

If both aggregation functions are sums and we the set of use all paths ${\cal A}$ while letting $m_{max} \rightarrow \infty$ then
\begin{equation}
    H_{\sum, \sum}(a; {\cal A}) =  \frac{1}{\alpha} \sum_m \sum_{j \in {\cal A}^m(a)} \alpha^m h(j) = \frac{1}{\alpha} \sum_m \sum_{j} (A^m)_{ja} \alpha^m h(j)
\end{equation}
is proportional to Alpha-Centrality. If we use $P$, the adjacency matrix weighted by out degree, instead of $A$ we get PageRank. As discussed, these metrics have limitations in the context of ESG rating, but it is interesting to see how they can be recovered in our framework.

The simplest new metric would be to take both aggregation functions equal to $\max$
\begin{equation}
    H_{MAX, MAX}(a; {\cal P}) = MAX( \{ \alpha^{m-1} x_{MAX}^m(a; {\cal P}) : 1 \leq m \leq m_{max} \})
\end{equation}
If $\alpha = 1$ the different ways of path counting are equivalent and this metric gives company $a$ a network ESG rating equal to the rating of the worst supplier in its network. If $\alpha < 1$ each supplier's harm is reduced by a factor $\alpha$ for every step from the target and the result is the worst weighted harm. More interesting are mixed aggregations, such as
\begin{equation}
    H_{MAX, AVG}(a;  {\cal P}) = MAX( \{ \alpha^{m-1} x_{AVG}^m(a;  {\cal P}) : 1 \leq m \leq m_{max} \} )
\end{equation}
which sets the network score of $a$ equal to the average harm of the worst level in the network, again with a factor $\alpha$ on the edges. This could, for example, highlight unethical or unsustainable practices at the base of the supply chain. 
\begin{equation}
    H_{AVG, MAX}(a;  {\cal P}) = \frac{ \sum_{m=1}^{m_{max}} \alpha^{m-1} x_{MAX}^m(a;  {\cal P}) }{\sum_{m=1}^{m_{max}} \alpha^{m-1} }
\end{equation}
measures the weighted average harm of the worst companies at each level and 
\begin{equation}
    H_{AVG, AVG}(a;  {\cal P}) = \frac{ \sum_{m=1}^{m_{max}} \alpha^{m-1} x_{AVG}^m(a;  {\cal P}) }{\sum_{m=1}^{m_{max}} \alpha^{m-1} }
\end{equation}
computes a weighted average harm over all nodes in the supply network. 

We could also use TOP-$k$ or any other aggregation function whatsoever.  In Section \ref{sec:small} we will analyse these four cases in detail since max and average represent something like polar opposite views on how to combine harm.

\subsection{Influence and Vulnerability}

With a procedure to determine network harm, $H(a)$, we can look at pairwise relationships, answering the question `how does $b$'s behaviour affect $a$'s score?' There are many ways to quantify this, the most direct being to see what the effect bad behaviour by $b$ has on $a$'s score. We call this \textbf{vulnerability}
\begin{equation}\label{eqn:vul}
    V_{AGG', AGG}(a,b) = H_{AGG', AGG}(a; {\cal P}, h(b)=100 ) -  H_{AGG', AGG}(a; {\cal P}) 
\end{equation}
Where $V$ will be a number between $0$ and $100$ representing how much worse the score of $a$ could get based on harmful actions by $b$.

Another option is to measure \textbf{influence}, which compares the network harm of $a$ to the value computed when $b$ is removed from the network. Define 
\begin{equation}\label{eqn:infl}
    I_{AGG', AGG}(a,b) = H_{AGG', AGG}(a; {\cal P},G-b) - H_{AGG', AGG}(a; {\cal P},G)
\end{equation}
where the notation $G-b$ means the network $G$ without the node $b$ or any of its connections. $I$ will be a number in the range $[-100,100]$, the value indicating how much the network harm would change by removing $b$. Note this does not simply mean that the target company cuts ties with $b$, since in that case $b$ could still contribute to $H(a)$ through second order and higher effects. Rather this represents completely removing $b$ from the network, say if it went out of business. 

\section{Results on Small Networks}\label{sec:small}
\begin{figure}
    \centering
    \includegraphics[width=1.1\linewidth]{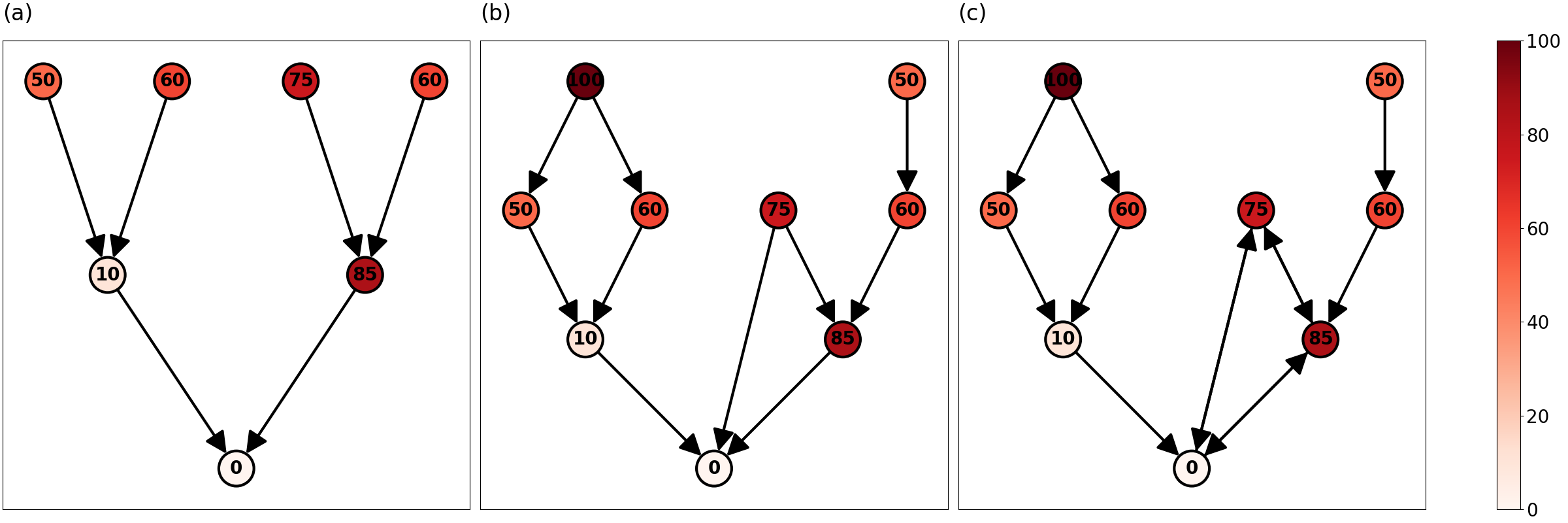}
    \caption{(a) Tree-like network. (b) Same network as (a) with two companies added at the base of the supply chain and a direct link between the company labelled 75 and the target. (c) Adds loops. Colour corresponds to harm, more red is means higher $h(i)$, labels show the exact values.}
    \label{fig:5}
\end{figure}
Starting with a simple example, Figure \ref{fig:5}(a) shows a tree-like supply network. In this case there is no difference between counting simple or shortest paths. With the bottom company (the only node labelled $0$) as the company of interest, the upstream harm values are 
\begin{align*}
x_{MAX} &= (x_{MAX}^1, x_{MAX}^2) = (85, 75)\\
x_{AVG} &= (x_{AVG}^1, x_{AVG}^2) = (47.5, 61.25)\\
\end{align*}
Using $\alpha=1$ for simplicity, the network harm values are (rounded to the nearest integer)
\begin{align*}
    H_{MAX, MAX} &= 85\\
    H_{MAX, AVG} &= 61\\
    H_{AVG, MAX} &= 80\\
    H_{AVG, AVG} &= 54
\end{align*}
$H_{MAX, MAX}$ measures the worst actor in the network; $H_{MAX, AVG}$ measures the worst layer; $H_{AVG, MAX}$ is the average of the worst actors in each layer and $H_{AVG, AVG}$ averages all the harm values in the network. 

The network in Figure \ref{fig:5}(b) adds more complexity. Now there are two paths from the node labelled 100 and two paths from the node labelled 75 and so the network harm depends on the method of path counting. 
\begin{align*}
x_{MAX} ({\cal C}) &= (85, 75, 100) \qquad x_{AVG} ({\cal C}) = (57, 61, 83)\\
x_{MAX} ({\cal S}) &= (85, 60, 100) \qquad x_{AVG} ({\cal S}) = (57, 57, 83)\\
x_{MAX} (\bar{\cal S}) &= (85, 60, 100) \qquad x_{AVG} (\bar{\cal S}) = (57, 57, 75)\\
\end{align*}
For example, in the shortest path version the worst node in level 2 has a score of $60$, since the node with $75$ is already considered at level 1. Hence the max value for that level is lower than with simple path counting, which considers the contribution of the node with $75$ in both layers. 

For the total harm, $H_{MAX, MAX}=100$ is the same regardless of the path counting approach. For the others, using $\alpha=1$ again, 
\begin{align*}
    H_{AVG, MAX}({\cal C}) &= 87\\
    H_{AVG, MAX}({\cal S}) &= 82\\
    H_{AVG, MAX}(\bar{\cal S}) &= 82
\end{align*}
with the difference explained by the difference in the upstream harms discussed above and
\begin{align*}
    H_{MAX, AVG}({\cal C}) &= 83\\
    H_{MAX, AVG}({\cal S}) &= 83\\
    H_{MAX, AVG}(\bar{\cal S}) &= 75
\end{align*}
with the difference in the single shortest path version due to counting the (very bad) 100 node only once. For completeness
\begin{align*}
    H_{AVG, AVG}({\cal C}) &= 67\\
    H_{AVG, AVG}({\cal S}) &= 66\\
    H_{AVG, AVG}(\bar{\cal S}) &= 63
\end{align*}
In a network like this, which tends to get worse as we go deeper, including more paths tends to increase the average harm.

Figure \ref{fig:5}(c) adds further complexity by introducing a number of new edges which create longer paths e.g. between the 85 node and the target there is now a direct link and path of length $2$. Using shortest paths gives the same result as before but using simple paths generates longer paths which particularly affect averages, pushing them towards node values which originate long paths. Including the path counting argument for clarity
\begin{align*}
x_{MAX} ({\cal C}) &= (85, 85, 100, 50)\\
H_{AVG, MAX}({\cal C}) &= 80\\
\end{align*}
This total harm is reduced because of the single, relatively low scoring, node that originates the only path of length $4$, $50\rightarrow60\rightarrow85\rightarrow75\rightarrow0$, in the network. This is undesirable behaviour. To mitigate it we can restrict the number of layers included, $m_{max}$, or we can use TOP-$k$ for the outer aggregation:
\begin{align*}
x_{MAX} ({\cal C}) &= (85, 85, 100, 50)\\
H_{TOP-50, MAX}({\cal C}) &= 90\\
\end{align*}
which averages only the worst layers. With values of $\alpha \ll 1$ this is less of an issue, as longer paths are suppressed by large powers of $\alpha$.

\begin{figure}
    \centering
    \includegraphics[width=0.95\linewidth]{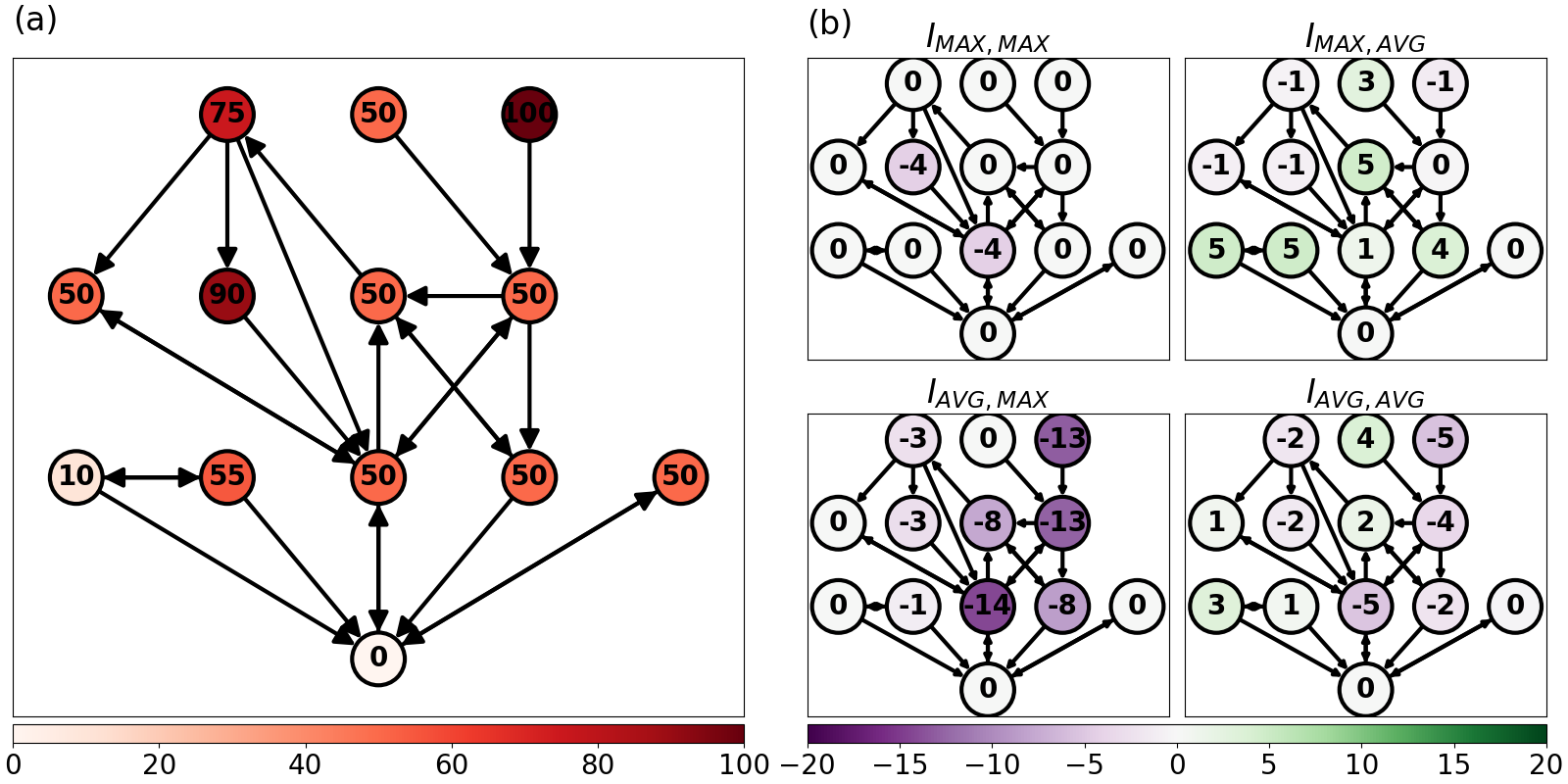}
    \caption{(a) A toy supply network for the bottom node, to a depth of 3. The longest simple path in this network is of length 7. (b) The  network coloured for influence on the bottom node, using different definitions of network harm and $\alpha = 0.85$ and counting all simple paths. }
    \label{fig:6}
\end{figure}
\begin{table}
    \centering
    \begin{tabular}{|c|c|c|c|c|}
    \hline
    Network Harm & $\alpha=0.85$ & $\alpha=0.15$ \\ \hline
    $H_{MAX, MAX}$    & 77 & 55  \\ \hline
    $H_{MAX, AVG}$    & 47 & 43 \\ \hline
    $H_{AVG, MAX}$    & 88 & 60 \\ \hline
    $H_{AVG, AVG}$    & 59 & 45 \\ \hline
    \end{tabular}
    \caption{ (a) Network harm values for high and low values of $\alpha$, for the network of Figure \ref{fig:5} rounded to the nearest integer, counting all simple paths ${\cal C}$. Counting shortest paths ${\cal S}$ gives similar results. }
    \label{tab:1}
\end{table}

Figure \ref{fig:6}(a) shows another (fictional) supply network with the target node at the bottom again, labelled 0. We compare harm metrics using $\alpha = 0.85$ and $\alpha = 0.15$ in Table \ref{tab:1}. $\alpha \simeq 0$ measures \textbf{intention}, a company is responsible for its direct suppliers, which have a factor $\alpha^0=1$, but is less responsible for the suppliers of its suppliers. $\alpha \simeq 1$ measures \textbf{consequence}, if a company's output is based on input with bad ESG characteristics, then it is penalised for this. In Table \ref{fig:1} the suppression of longer paths for $\alpha = 0.15$ means only companies close to the target contribute to the score.

The analysis is more interesting, and most resistant to `greenwashing' or obfuscation, if we use a large value of $\alpha$. We do this to produce Figure \ref{fig:6}(b) which measures influence. Different network harm metrics highlight different companies. For $H_{MAX, MAX}$ note that the node with the worst possible rating, 100, is three steps from the target since $\alpha^3 = (0.85)^3 \simeq 0.61$ the closer bad nodes, in particular the one rated 90, are more influential. The central 50 node in the first layer which connects the target to the 90 node is also quite influential. The mixed versions, $I_{AVG, MAX}$ and $I_{MAX, AVG}$, enhance this effect and seem particularly apt for detecting harm propagating through the supply chain.

\section{Results on Real Networks}\label{sec:realdata}

\subsection{Company Network}\label{sec:ekodata}

We obtained ESG ratings and company interaction data from ekoIntelligence \footnote{\url{https://www.ekointelligence.com/}}, an ESG rating agent. They use a mixture of public and private information to develop ESG ratings for companies. As with all ESG raters, their exact methods are a trade secret. While they provide detailed breakdowns of each company they rate, we only required the ratings which are given as values between 0 and 100 and represent preliminary analysis of each company's ESG performance. We converted these values to harm scores using $(100-\text{ekoIntelligence score})$.

\begin{figure}[htbp]
    \centering
    \includegraphics[width=0.9\textwidth]{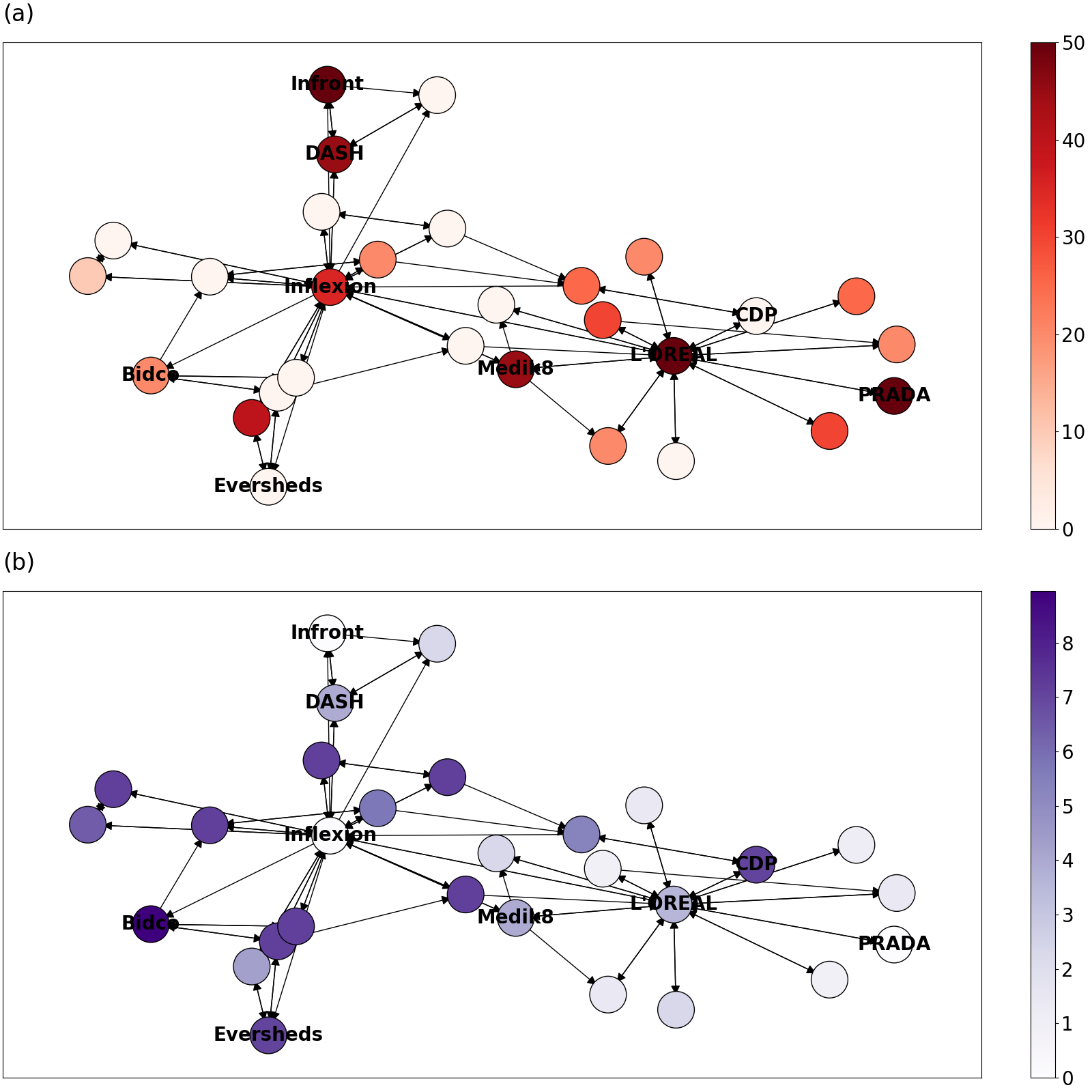}
    \caption{(a) The company network, with harm scores indicated by the colour. (b) The vulnerability of the target company (INFLEXION) to bad behaviour of suppliers using $H_{MAX, AVG}({\cal S})$ and $\alpha=0.85$. }
    \label{fig:eko}
\end{figure}
The dataset was constructed by snowball sampling from the central company Inflexion. As the network was expanded outward from this seed, the aim was to gather connections between companies in the sample, but not every relationship of every company. Our purpose here is to illustrate how the proposed methods operate on realistic company ESG data, rather than to make claims about the ESG performance of Inflexion or any particular company. As such, although many relationships with other companies were identified we focus on the core where data is more accurate\footnote{This was identified by first computing the 5-core of the whole company network, then merging entities which were different `branches' of the same company e.g. L'Oréal S.A., L'Oréal U.K. etc.}. The network and its harms are shown in Figure \ref{fig:eko}(a). Of perhaps greatest interest to investors and the target company themselves is Figure \ref{fig:eko}(b) where we show the vulnerability of the target to bad behaviour in its supply chain, given a particular definition of network harm $H_{MAX, AVG}({\cal S})$ and $\alpha=0.85$. The top 3 companies Inflexion is most vulnerable to potential harmful behaviour by are Bidco, Eversheds and CDP, because they contribute to multiple paths and have low intrinsic harm, so currently do not contribute much to Inflexion's harm score.

\subsection{International Trade Network}

Countries may also be concerned with the ESG rating of their trading partners and with international trade data we can gather a more complete network which allows more detailed analysis. International trade data is available from CEPII \cite{mayer2023cepii}. This database describes trade flows between 162 countries in 9 industrial sectors. To construct a trade network we take data from 2020 and sum the total trade across all 9 sectors between every pair of countries. We draw an unweighted, directed link between two countries if the total trade flow from $a$ to $b$ exceeds 100 million USD.

Country level ESG data can be obtained from the World Bank \cite{WorldBankESG}. They provide 71 indicators across 239 countries. There are a wide variety of indicators covering everything from carbon emissions to the proportion of female politicians. Since the aim of this paper is to illustrate network harm analysis, we focus on a subset of environmental indicators: \textit{$CO_2$ emissions, Level of water stress, Methane emissions, Nitrous oxide emissions, Renewable energy consumption, PM2.5 air pollution} which are easy to interpret. To convert these values to harm scores, for each indicator $x$ we compute a score for country $c$ as
\begin{equation}
    \bar{x}_c = 100 \times \frac{x_c - \min(x)}{\max(x) - \min(x)}
\end{equation}
where the max and min are taken across all countries for which data is available and $x_c$ is the original value for $c$ obtained from the World Bank data. For $x = $\emph{Renewable energy consumption}, where more is better, we convert it to a harm score in the usual way, $100 - x$. We exclude countries where all 6 scores are not available and compute the intrinsic harm $h(c)$ as the average of the top 3 worst indicator scores. This gives greater range in scores than a simple average, which is useful for differentiating countries. Again, the point here is not that this is an ideal way to aggregate ESG indicators, rather it is to construct `reasonable' harms scores and study how they propagate on the trade network.

\begin{figure}
    \centering
    \includegraphics[width=0.95\linewidth]{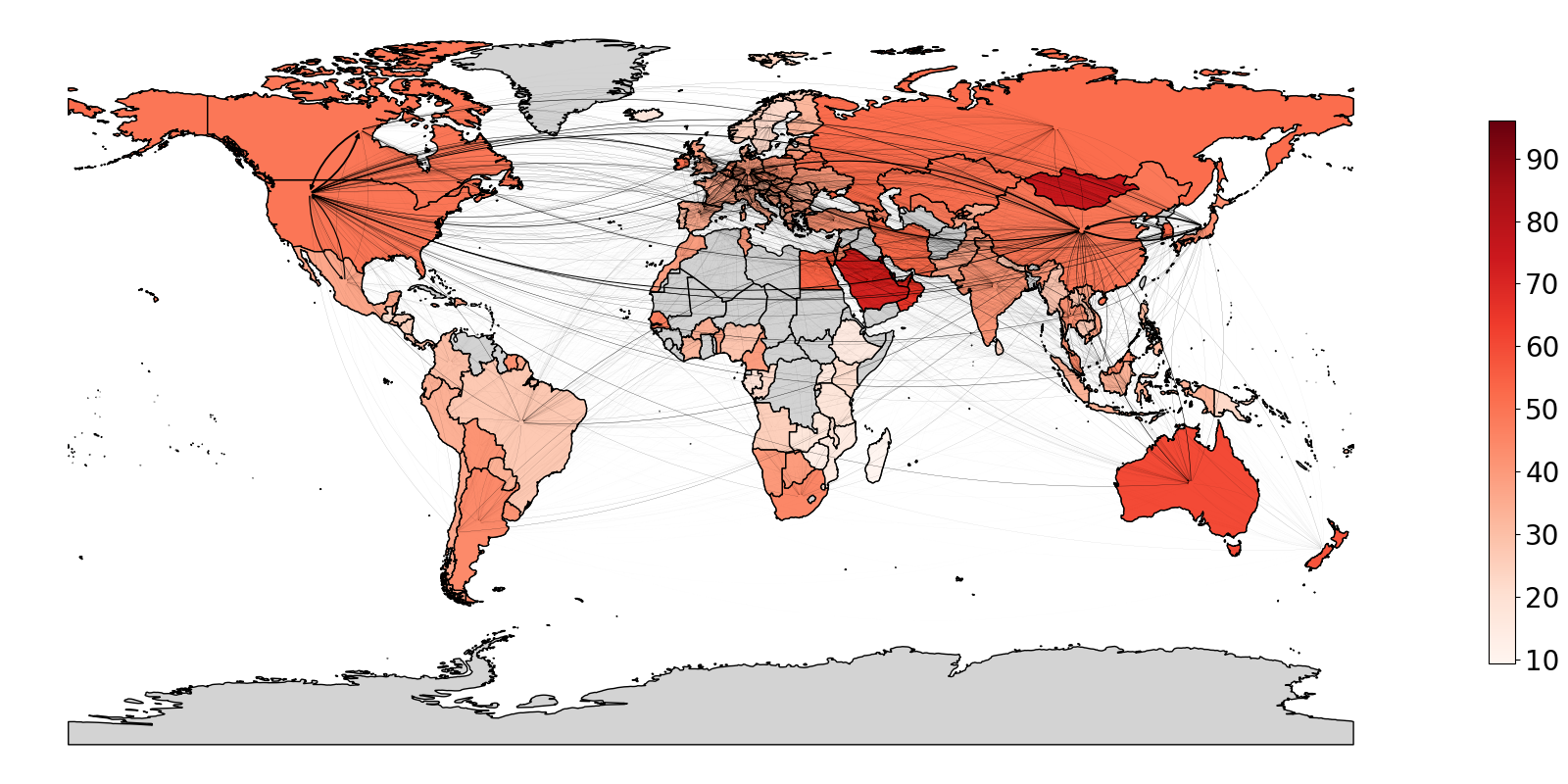}
    \caption{The trade flow network constructed from CEPII and WorldBank data as described in the text. Colour indicates harm, more red is worse. }
    \label{fig:map}
\end{figure}
\begin{figure}
    \centering
    \includegraphics[width=0.95\linewidth]{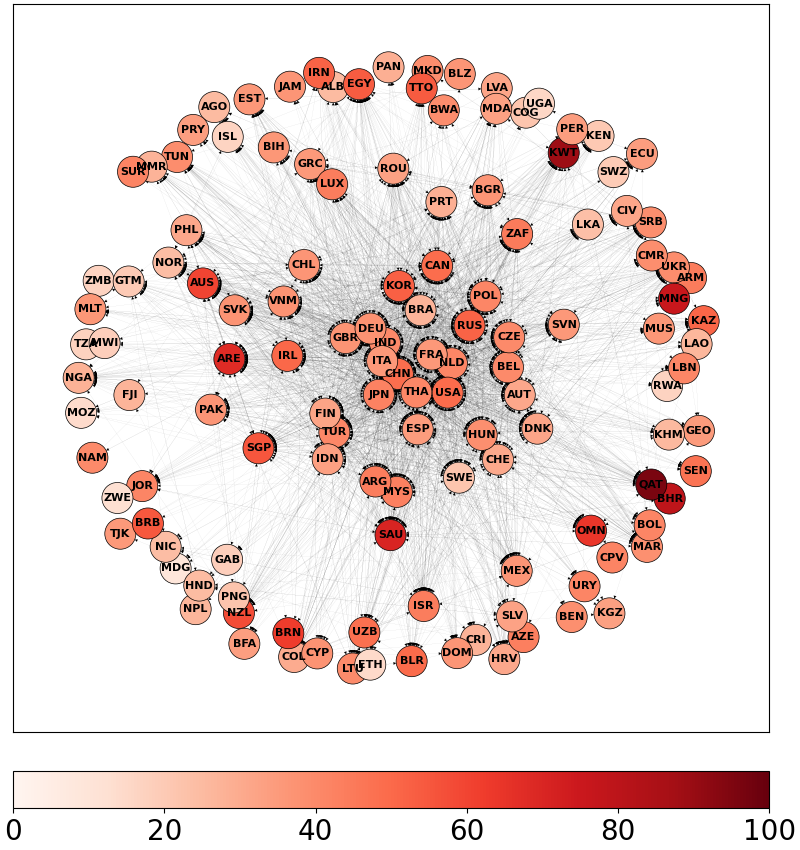}
    \caption{Topological view of the country data from Figure \ref{fig:map}. }
    \label{fig:7}
\end{figure}
\begin{table}
    \centering
    \begin{tabular}{|c|c|c|c|}
    \hline
    \multicolumn{2}{|c|}{ Bottom 5 Intrinsic Harm } & \multicolumn{2}{|c|}{Top 5 Intrinsic Harm } \\
    \hline
    Country & $h$ &  Country & $h$ \\ \hline
Madagascar (MDG) & 9.3 & Qatar (QAT) & 96.1 \\
Zimbabwe (ZWE) & 12.6 & Kuwait (KWT) & 88.9 \\
Mozambique (MOZ) & 14.3 & Bahrain (BHR) & 79.0 \\
Ethiopia (ETH) & 14.7 & Mongolia (MNG) & 75.9 \\
Uganda (UGA) & 15.8 & Saudi Arabia (SAU) & 70.6 \\ \hline
    \end{tabular}
    \caption{ Highest and lowest intrinsic harm scores computed as described in the text. Mongolia has a high score because of high levels of $NO_2$ and very low renewable usage. The others high scoring countries are oil producing gulf states. }
    \label{tab:2}
\end{table}
We construct the final trade flow network by removing any country without a valid harm score, any country with no incoming links and any isolated nodes, leaving 131 countries. The network is shown in Figures \ref{fig:map} and \ref{fig:7}. The highest and lowest harm scores are shown in Table \ref{tab:2}. For this network, computing all simple paths is time consuming and unnecessary, since countries are much more likely to enter into bilateral relationships than companies so almost every pair is connected by a path of length at most 2. Thus we use all shortest paths and for aggregation we use $H_{MAX, AVG}$ and $\alpha = 0.85$.

\begin{figure}
    \centering
    \includegraphics[width=0.95\linewidth]{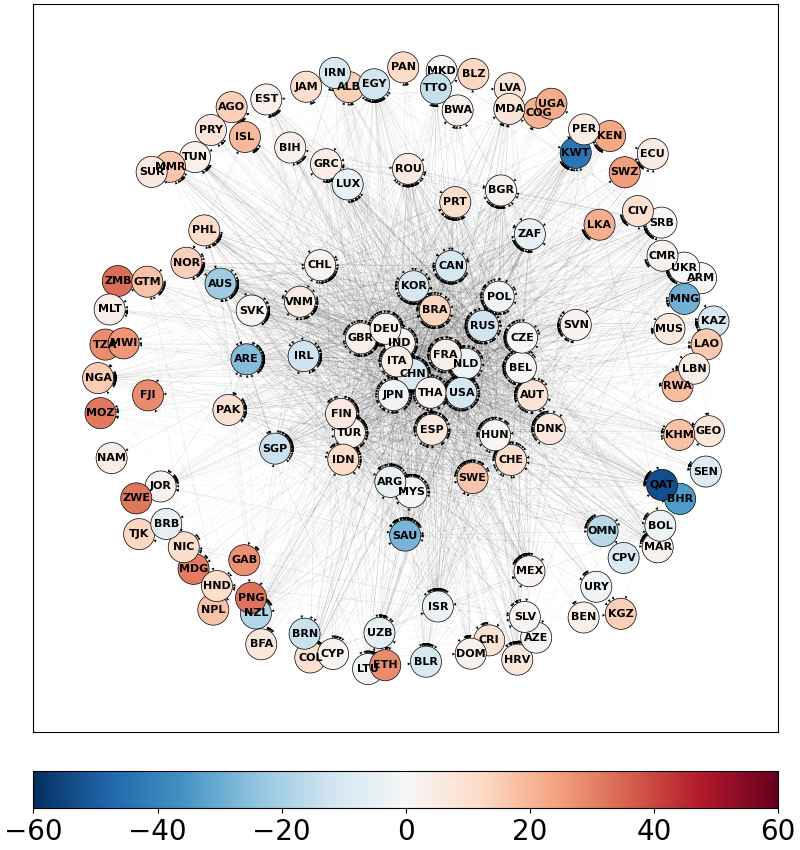}
    \caption{The trade flow network. Colour indicates $H_{MAX, AVG}(c) - h(c)$, comparing the network harm to the intrinsic harm. }
    \label{fig:8}
\end{figure}
Figure \ref{fig:8} shows the difference $H_{MAX,AVG}(c) - h(c)$, which highlights countries whose network harm is worse than their intrinsic harm. One important effect to note here, because of the dense network, short path lengths and because the intrinsic score of a country doesn't influence its own network score, the countries with the worst intrinsic harm tend to have lower network harm and vice versa, because they don't `pollute' their own network.

\begin{figure}
    \centering
    \includegraphics[height=0.95\textheight]{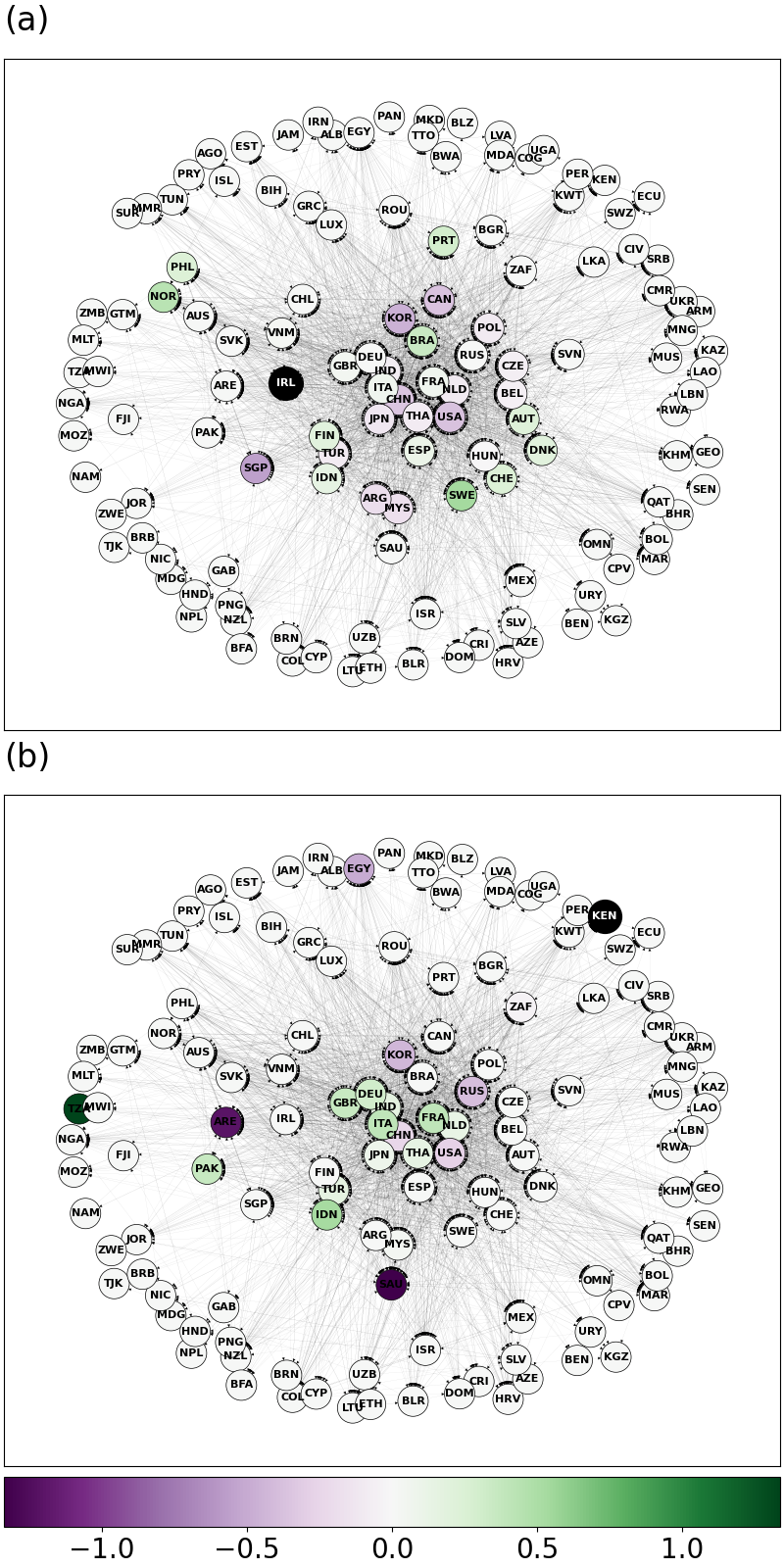}
    \caption{The trade flow network where colour indicates influence score. (a) Uses Ireland (IRL) as the target and (b) uses Kenya (KEN). }
    \label{fig:9}
\end{figure}
Figure \ref{fig:9} shows influence scores, $I_{MAX, AVG}$, for two example countries, Ireland and Kenya. Ireland has average to high intrinsic harm (50.5) and lower network harm (38). Its most harmful direct supplier is Kuwait (88.9) but the country which has the most negative influence on its score is Singapore, since it is a conduit for many paths to Ireland. Ireland's network score is comparatively low because it trades with many European countries with lower scores than it. Kenya has quite low intrinsic harm (20.7) and higher network harm (43.9). Its worst direct supplier is the USA (49.4) but the countries which most influence its network harm are Saudi Arabia and the United Arab Emirates, which are not direct trading partners. Its score is raised because its direct partners \emph{are} directly supplied by these high harm nodes.

\begin{table}
    \centering
    \begin{tabular}{|c|c|c|c|}
    \hline
    \multicolumn{2}{|c|}{ Largest Negative GI } & \multicolumn{2}{|c|}{Largest Positive GI} \\
    \hline
    Country & $GI_{MAX, AVG}$ &  Country & $GI_{MAX, AVG}$ \\ \hline
United States (USA) &   -146 & Brazil (BRA) &  36 \\
South Africa (ZAF) &   -90 & France (FRA) &   33 \\
China (CHN) &   -69 & Sweden (SWE) &   31 \\
Saudia Arabia (SAU) &  -34 & Italy (ITA) &   21 \\
United Arab Emirates (ARE) &  -34 & Switzerland (CHE) &   20 \\ \hline
    \end{tabular}
    \caption{ Highest and lowest global influence (GI) scores computed as described in the text. See text for discussion. }
    \label{tab:3}
\end{table}
\begin{figure}
    \centering    
    \includegraphics[width=0.95\linewidth]{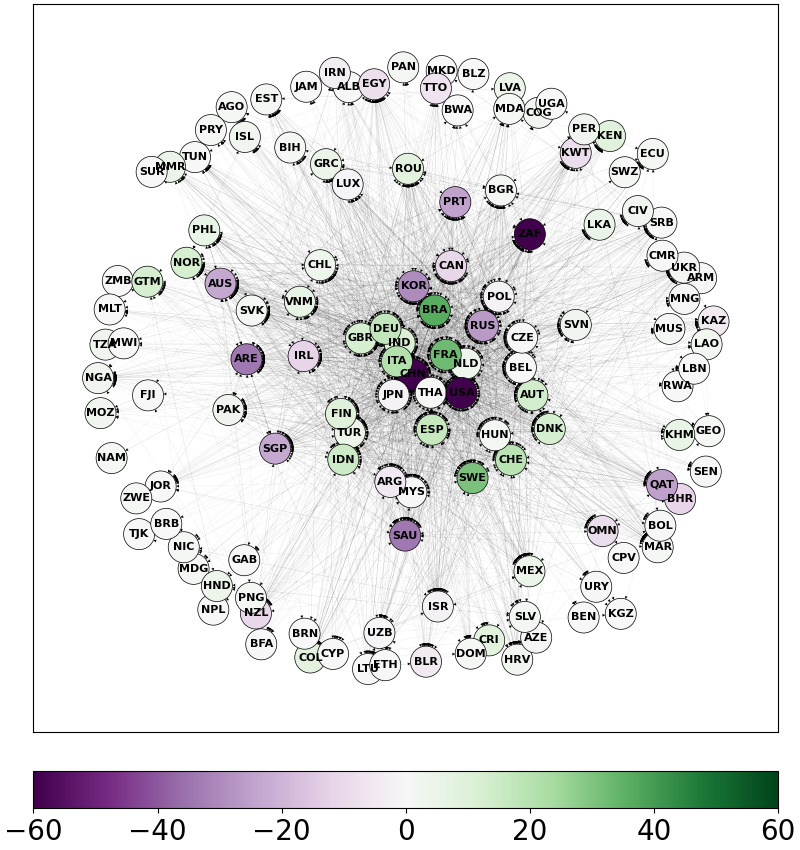}
    \caption{The global influence score $GI_{MAX, AVG}$ as defined in the text, computed on the trade flow network. }
    \label{fig:10}
\end{figure}
Finally, we can aggregate influence scores to rank countries based on how their own ESG ratings, their connections and position in the network affect others. We call this score \emph{global influence} which we compute as
\begin{equation}
    GI_{AGG', AGG}(a) = \sum_{n \neq a} I_{AGG', AGG}(n, a)
\end{equation}
that is $GI(a)$ is the sum of the influence $a$ has on every other country. The highest and lowest $GI_{MAX, AVG}$ scores are shown in Table \ref{tab:3} and all the scores are visualised in Figure \ref{fig:10}. Countries with large negative $GI$ are either themselves badly rated (Saudia Arabia) or are directly connected to many badly rated countries (United States). Large positive scores tend to be  associated with European countries (though Brazil is highest) which have low to moderate harms themselves and trade mostly with similarly rated countries. Peripheral countries with good to moderate ratings have little influence in either direction.

\section{Conclusion}\label{sec:conclusion}

In this paper we have introduced a family of metrics for evaluating `harm' in the interaction network of some entity. Rather than proposing a one-size-fits-all approach, we recognise the diversity of opinions and approaches to ESG rating. Our framework is quite flexible with respect to the way harms are aggregated and the way we deal with complex dependencies through path counting. We have analysed synthetic and real networks to show the use of these metrics in practice. They largely behave as expected and can be used to highlight various aspects of ESG `pollution' in the interaction network. By tuning a parameter, $\alpha$, and the method of path counting we can develop metrics that are more robust to the greenwashing motifs discussed in the Introduction. Probably most interestingly we can study the effect different entities in the network have on the scores of some target. This can allow decision makers within companies and nations to make conscious and tactical decisions about who they trade with based on how that will change their own ESG rating.

If bad ESG ratings can `pollute' the rest of the network, the hope is this will encourage peer pressure or isolation of offending companies, realising the goals of ESG investing to encourage responsible corporate behaviour. While greenwashing and gaming of metrics is a possibility, we have provided a flexible framework where the method of analysis can be tuned to avoid intentional or unintentional greenwashing which hides bad behaviour deeper in the network. The most obvious generalisation of this approach is to allow for weighted edges, measuring e.g. the volume of goods traded. We leave this for future work since it is not obvious how trade volume influences ESG harm propagation e.g. the difference in influence between 1 and 2 billion dollars of trade is not the same as the difference between 1 dollar and 1 billion dollars!

While we are primarily concerned with ESG rating, and have developed our methods for this purpose, the framework proposed is general enough to be used for any application where we want to accumulate values or compute centrality on a node valued network. As already mentioned, carbon accounting in Life Cycle Analysis \cite{guinee2011life} is one potential application using sum aggregation. Other examples where this framework may be useful include epidemic spreading \cite{colizza2006role}, social media \cite{arthur2019human} or even protein-protein interaction \cite{tosadori2021analysing}. We leave these and other extensions for future work.

\section*{Acknowledgements} 
This work was supported by funding for the AI for Collective Intelligence Research Hub from the UKRI AI Programme and EPSRC (grant ref EP/Y028392/1). We thank Richard Nicholson and Neil Ellis for fruitful discussion and help with the ekoIntelligence data.

\bibliographystyle{plain}
\bibliography{references}

\end{document}